\documentstyle[11pt,newpasp,twoside,epsf]{article}
\def\edcomment#1{\iffalse\marginpar{\raggedright\sl#1\/}\else\relax\fi}
\reversemarginpar

\begin{document}
\title{Supermassive Black Holes in Disk Galaxies}
\author{John Kormendy}
\affil{Department of Astronomy, RLM 15.308, University of Texas,
      Austin, TX 78712, USA (kormendy@astro.as.utexas.edu)}

\begin{abstract}
Dynamical searches find central dark objects -- candidate supermassive
black holes (BHs) -- in at least 34 galaxies. The demographics of these objects
lead to the following conclusions: 

(1) BH mass correlates with the luminosity of the bulge component of the
    host galaxy, albeit with considerable scatter.  The median BH mass fraction
    is 0.15\thinspace\% of the mass of the bulge.  The quartiles are 
    0.09\thinspace\% and 0.7\thinspace\%.

(2) BH mass correlates with the mean velocity dispersion of the bulge 
    component inside its effective radius, i.{\thinspace}e., with how strongly
    the bulge stars are gravitationally bound to each other.  For the best BH
    mass determinations, the scatter is consistent with the measurement errors.

(3) BH mass correlates with the luminosity of the high-density central
    component in disk galaxies independent of whether this is a real bulge 
    (a mini-elliptical) or a ``pseudobulge'' (believed to form via inward
    transport of disk material). 

(4) BH mass does not correlate with the luminosity of galaxy disks.  If pure
    disks contain BHs (and AGN observations argue that some do), then their BH
    mass fractions are much smaller than the canonical 0.15\thinspace\% for
    bulges.

These results lead to the following conclusions:

(A) Present observations show no dependence of BH mass on the details of whether
    BH feeding happens rapidly during a collapse or slowly via secular evolution
    of the disk.

(B) The above results increasingly support the hypothesis that the major events
    that form a bulge or elliptical galaxy and the main growth phases of its
    BH -- when it shone as an AGN -- were the same events. 
\end{abstract}

\vfill\eject

\section{Introduction}

\lineskip=0pt \lineskiplimit=0pt
\pretolerance=15000  \tolerance=15000

\newdimen\sa  \def\sd{\sa=.1em  \ifmmode $\rlap{.}$''$\kern -\sa$
                                \else \rlap{.}$''$\kern -\sa\fi}

      A major payoff period for the {\it Hubble Space Telescope\/} (HST) is
under way, as the Space Telescope Imaging Spectrograph (STIS) starts to produce
black hole detections routinely and in large numbers.  Eight new BH detections
are available from Nuker team work (Gebhardt et nuk.\ 2000b), and improved
observations or new detections are available for 6 more
(Verdoes Kleijn et al.\ 2000; Bower et al.\ 2000; Green et al.\ 2000; Kaiser et
al.\ 2000; Nelson et al.\ 2000).  The total number of BHs available for
demographic studies is now at least 34.  This is enough for a quantum
improvement in our ability to ask astrophysical questions.

     STIS is well designed for black hole searches based on stellar or ionized
gas dynamics.  In addition, maser disks have provided BH detections in three
galaxies, including the spectacular case of NGC 4258 (Miyoshi et al.\ 1995).
What dynamical searches securely detect are central dark objects with masses
\hbox{$M_\bullet$ $\sim$ 10$^6$ -- 10$^{9.5}$ $M_\odot$.}  The arguments that
these dark masses are BHs and not dark clusters of (e.{\thinspace}g.)~brown
dwarf stars, white dwarf stars, neutron stars, or stellar-mass black holes are
indirect.  However, in NGC 4258 and in our Galaxy, the maximum possible radius
that a dark cluster can have is so small that astrophysical constraints
(collision times for brown dwarfs and evaporation times for stellar remnants)
rule out the above BH alternatives (Maoz 1998).  This result increases our
confidence that the other central dark objects also are BHs.

      With STIS, the BH search has become feasible for most nearby galaxies 
that have unobscured centers and reasonably old stellar populations.  The search
is not easy; e.{\thinspace}g., searching for a 10$^6$-$M_\odot$ BH is still
difficult at the distance of the Virgo cluster and impossible much beyond.  
But well chosen samples provide reliable detections or upper limits on 
$M_\bullet$ that are suitable to address a variety of astrophysical questions.
Many observing programs of this sort are currently under way and are, as of
mid-2000, just beginning to yield results.

      My oral paper reviewed four particularly good BH cases, NGC 4258 (the
maser galaxy), NGC 4374 (which contains an ionized gas disk with a very clean 
BH signature -- Bower et al.\ 1998), NGC 3115 (whose nucleus,
with an observed dispersion of 600 $\pm$ 37 km s$^{-1}$, would be unbound
without a dominant dark object -- Kormendy et nuk.\ 1996), and our Galaxy.  The
latter is now the strongest BH case, based on measurements of stellar proper
motions in a star cluster within $\sim 0\sd5$ = 0.02 pc of Sgr
A* (Eckart \& Genzel 1997; Genzel et al.\ 1997, 2000; Ghez et al.\ 1998).  The
fastest-moving star has a total proper motion of $1600 \pm 200$ km s$^{-1}$.
Genzel and Ghez -- and we, as spectators -- can look forward to observing the
Galactic center rotate once in our lifetimes!  Already preliminary accelerations
have been measured (Ghez et al.\ 2000). One reason why this is important is that
the three available acceleration vectors do in fact intersect near Sgr A*.  

      In this written version of my paper, I concentrate on results provided by
the new BH detections.

\vfill\eject

\section{The $M_\bullet$ --
         $M_{B,{\rm\char'142\char'165\char'154\char'147\char'145}}$ and
         $M_\bullet$ -- $\sigma_e$ Correlations}

      Two fundamental correlations have emerged from BH detections.  Figure 1
(left) shows the correlation between BH mass and the luminosity of the ``bulge''
part of the host galaxy (Kormendy 1993a, Kormendy \& Richstone 1995; Magorrian
et nuk.\ 1998) brought up to date with new detections.  A least-squares fit 
gives 
\begin{equation}
M_{\bullet} = 0.93\times10^8~M_\odot\left(L_{B,\rm
bulge}\over{\hbox{$10^{10}~L_{B\odot}$}}\right)^{1.11}.
\end{equation}
Since $M/L \propto L^{0.2}$, Equation (1) implies that BH mass is, within
errors, proportional to bulge mass: $M_\bullet \propto M_{\rm bulge}^{0.93}$. 

      Figure 1 (right) shows a newly discovered correlation between BH mass and
the luminosity-weighted velocity dispersion $\sigma_e$ within the effective
radius $r_e$ (Gebhardt et nuk.\ 2000a; Ferrarese \& Merritt 2000).  The line is a
fit to the data assuming that errors in $\sigma_e$ are zero and that errors in
$\log$\,$M_\bullet$ are the same for each galaxy.  A least-squares fit to the
subset of galaxies with most reliable $M_\bullet$ measurements (see Gebhardt
et nuk.\ 2000a) gives
\begin{equation}
M_{\bullet}
=1.2\times10^8~M_\odot\left(\sigma_e\over\hbox{200~{\rm km~s$^{-1}$}}\right)^{3.75}.
\end{equation}

\vfill

\includegraphics{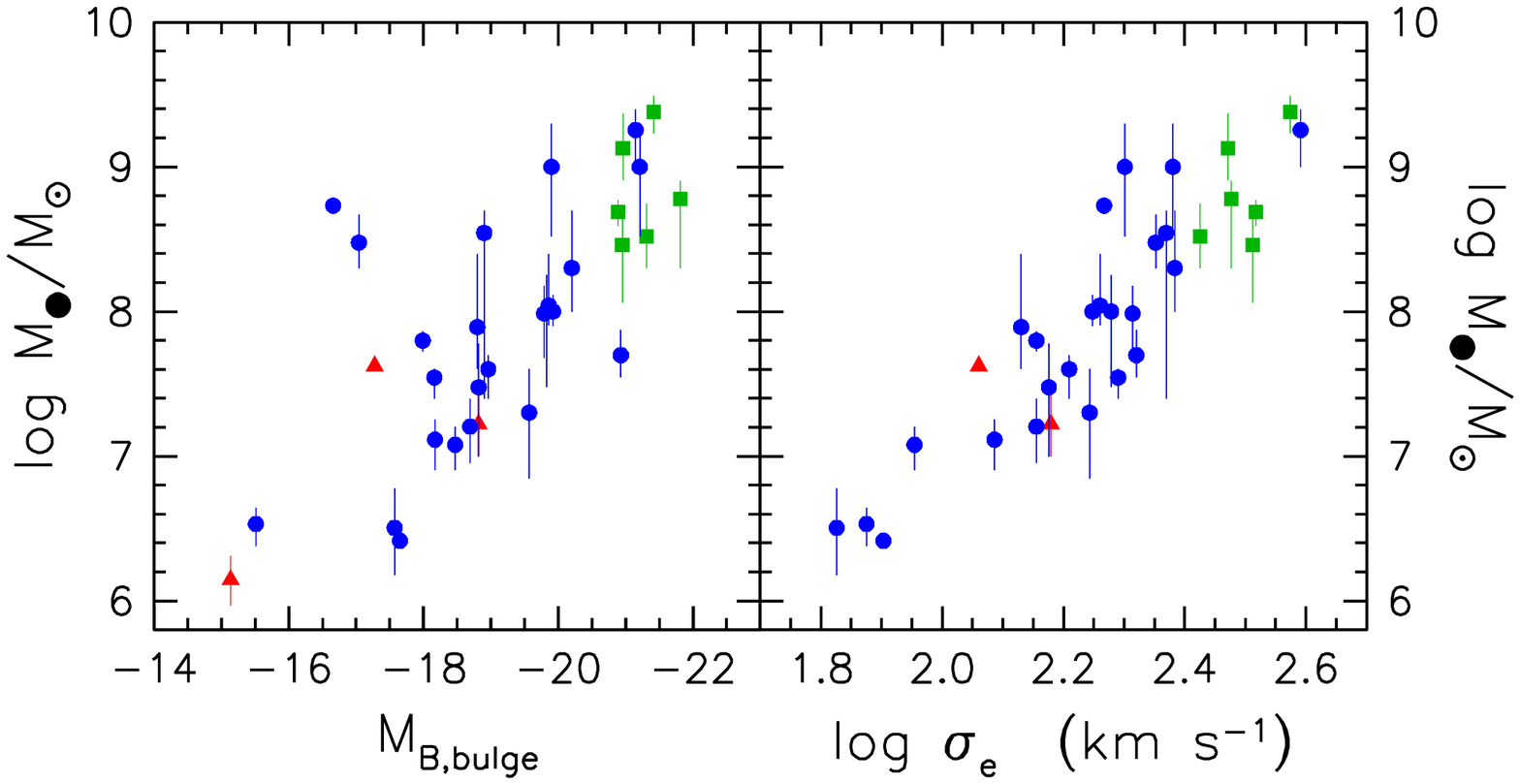}
\vskip 10pt
\noindent Figure 1.{\hskip 1.5em}(left) Correlation of BH mass with the absolute
magnitude of the bulge component of the host galaxy.  (right) Correlation of BH
mass with the luminosity-weighted mean velocity dispersion inside the effective
radius of the bulge.  In both panels, filled circles indicate $M_\bullet$
measurements based on stellar dynamics, squares are based on ionized gas
dynamics, and triangles are based on maser disk dynamics.  All three techniques
are consistent with the same correlations.  However, note that BH mass error
bars are derived within the assumptions of the models used, so the true errors
are underestimated for $M_\bullet$ values based on two-integral models.  This is
particularly relevant for the most discrepant galaxy at left, NGC 4486B.

\eject

      The scatter in the $M_\bullet$ -- $M_{B,\rm bulge}$ correlation is large:
the RMS dispersion is a factor of 2.8 and the total range of BH masses is about two orders of magnitude at a given $M_{B,\rm bulge}$.  There are also two exceptions with unusually high BH masses.  The most extreme case, NGC 4486B
(Kormendy et nuk.\ 1997; Green et al.\ 2000), is still based on two-integral
models.  Its BH mass may decrease when three-integral models are constructed.
Despite the scatter, the 34 detections imply a robust correlation.  Many authors
have wondered whether the apparent $M_\bullet$ -- $M_{B,\rm bulge}$ correlation
is real or only the upper envelope of a distribution that extends to smaller
$M_\bullet$.  The latter possibility is now unlikely: ongoing searches find BHs
in essentially every bulge observed, and the error bars in $M_\bullet$ are
small.  Put another way, almost all BHs represented in Figure 1 would have been
detected even they if were significantly farther away.

      In contrast, the scatter in the $M_\bullet$ -- $\sigma_e$ correlation is
small, and the galaxies that were discrepant above are not discrepant here.
Gebhardt et nuk.\ (2000a) show that the scatter is essentially consistent
with measurement errors for galaxies with the most reliable $M_\bullet$
measurements.  Therefore the $M_\bullet$ -- $\sigma_e$ correlation is
more fundamental than the $M_\bullet$ -- $M_{B,\rm bulge}$ correlation.  What
does this mean?  What does it tell us about galaxy formation and BH growth?

      Both correlations imply a close connection between BH growth and galaxy
formation.  They suggest that the BH mass is determined in part by the amount of
fuel that is available; this is connected with the total mass of the bulge.  

      But when was the mass accreted?  There are three generic possibilities.
(1) BHs could have grown to essentially their present masses before galaxies
formed and then regulated the amount of galaxy that grew around them
(e.{\thinspace}g., Silk \& Rees 1998).  (2) Seed BHs that formed early or that
were already present at the start of galaxy formation could have grown to their
present masses as part of the galaxy formation process.  This would mean that
the major merger and dissipative collapse events that made a bulge or elliptical
galaxy were the same events that made quasars shine (e.{\thinspace}g., Sanders
et al.\ 1988a, b).  (3) The majority of BH mass may have been accreted after
galaxy formation from ambient gas in the bulge.

      Figure 1 provides the first of two arguments (the other is in
\S{\thinspace}4) in favor of (2), that BHs and galaxies form together.
Exceptions to the $M_\bullet$ -- $M_{B,\rm bulge}$ correlation satisfy
the $M_\bullet$ -- $\sigma_e$ correlation.  This means that, whenever a BH is
unusually high in mass for a given luminosity, it is also high in
$\sigma_e$ for that luminosity.  That is, it is high in the Faber-Jackson
(1976) $\sigma(L)$ correlation.  One possible reason might be that the 
mass-to-light ratio of the stars is unusually high; this proves not to be
the main effect.  The main effect is illustrated in Figure 2.  Ellipticals that
have unusually high dispersions for their luminosities are unusually compact:
they have unusually high surface brightnesses and small effective radii for
their luminosities.  Similarly, unusually cold galaxies are unusually fluffy:
they have low effective surface brightnesses and large effective radii.  
Therefore, when a galaxy is observed to be hotter than average, we conclude
that it underwent more dissipation than average and shrunk inside its dark
halo to a smaller size and higher density than average.  We will say that it
``collapsed'' more than average.  {\it If BHs are unusually massive
whenever galaxies are unusually collapsed, then this strongly suggests that
BH masses were determined by the collapse process.\/} 

      Consider the alternative: Could unusually massive BHs {\it cause} a galaxy
to be unusually collapsed.  This seems unlikely, because more massive BHs would
power more luminous quasars.  Their radiation pressure would tend to push on 
the protogalactic gas and make it collapse less, not more, than average.

      The $M_\bullet$ -- $M_{B,\rm bulge}$ and $M_\bullet$ -- $\sigma_e$
correlations contain {\it almost\/} the same information.  After all, the 
mass of the bulge involves $\sigma_e^2$ and (small) corrections for velocity
anisotropy.  But the $M_\bullet$ -- $\sigma_e$ correlation contains something
new: it contains information about how much the bulge collapsed when it formed.
BH mass correlates better with this new information about the galaxy formation
process than it does with the amount of raw material involved. \hbox{Put another
way,} $M_{\rm bulge} \propto \sigma_e^2 r_e / G$; it is the presence of $r_e$
that makes a galaxy an exception to the $M_\bullet$ -- $M_{B,\rm bulge}$
correlation when it is not an exception to the $M_\bullet$ -- $\sigma_e$
correlation.  All this emphasizes the close connection between BH masses and 
the bulge formation process.

\vfill
      
\vskip 10pt

\includegraphics{collapse.ps}

\noindent Figure 2.{\hskip 1.5em}Correlations with absolute magnitude of velocity dispersion (upper
panel), effective surface brightness (middle panel) and effective radius
(lower panel) for elliptical galaxies from the Seven Samurai papers.
Galaxies with unusually high or low velocity dispersions for their
luminosities are identified in the top panel and followed in the other panels.

\eject

\section{The $M_\bullet$ -- $M_{B,{\rm total}}$ Correlation: BHs Do Not Know About Disks} 

\lineskip=0pt \lineskiplimit=0pt

      It is important to note that BH mass does not correlate with disks in the
same way that it does with bulges.  Figure 3 shows the correlations of BH mass
with the absolute magnitude of the bulge (left) and with the total
absolute magnitude of the galaxy (right).   \hbox{Figure 3} (right) shows that
disk galaxies with small bulge-to-total luminosity ratios destroy the reasonably
good correlation seen in Figure 3 (left).  In addition, Figure 3 (right) shows
four galaxies that have strong BH mass limits but no bulges.  They further
emphasize the conclusion that disks do not contain BHs with nearly the same mass
fraction as do bulges.  In particular, in the bulgeless galaxy M{\thinspace}33,
the upper limit on a BH mass from STIS spectroscopy is $M_\bullet$
$_<\atop{^\sim}$ 2000 $M_\odot$.  If M{\thinspace}33 contained a BH with the
median mass fraction observed for bulges, then we would expect that $M_\bullet
\sim 3 \times 10^7$ $M_\odot$.

      Figure 3 tells us that BH masses do not ``know about'' galaxy disks.
Rather, they correlate with the high-density bulge-like component in galaxies.

      The above results do not preclude BHs in pure disk galaxies as long as
they are small enough.  Filippenko \& Ho (2000) emphasize that some pure
disks are Seyfert galaxies.  They probably contain BHs.  The extreme example
is NGC 4395, the lowest-luminosity Seyfert known.  NGC 4395 is included in
Figure 3.  However, if its BH were radiating at the Eddington rate,
then its mass would be only $M_\bullet \sim 100$ $M_\odot$ (Filippenko
\& Ho 2000).  This is consistent with the dynamical mass limit.  So disks can
contain BHs, but their masses are {\it much\/} smaller in relation to their disk
luminosities than are bulge BHs in relation to bulge luminosities.  It is
possible that the small BHs in disks are similar to the seed BHs that once must
have existed in protobulges, too, before they grew monstrous during the AGN and
galaxy formation era.

\vfill
      
\vskip 10pt

\includegraphics{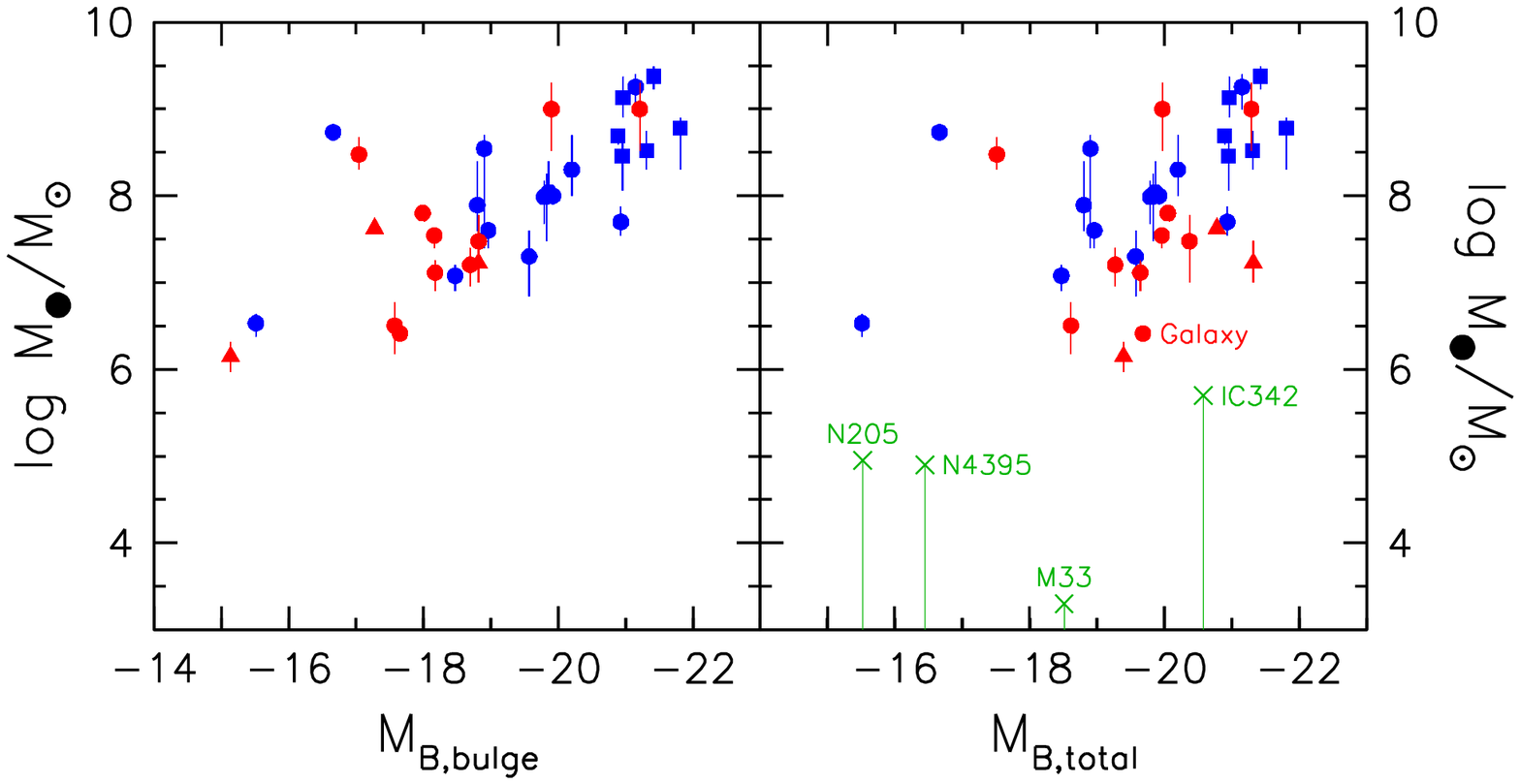}

\noindent Figure 3.{\hskip 1.5em}(left) $M_\bullet$ -- $M_{B,\rm bulge}$ correlation from Figure 1.  (right) Plot of $M_\bullet$ against the total
absolute magnitude of the host galaxy.   Blue points denote elliptical
galaxies, red points denote bulges of disk galaxies, and green crosses denote
galaxies that do not contain either a bulge or a pseudobulge (\S\thinspace4).
Spiral galaxy limits: M{\thinspace}33 (Scd) is from Gebhardt et al.\ (2000b);
IC 342 (Scd) is from B\"oker et al.\ (1999), and NGC 4395 (Sm) is from
Filippenko \& Ho (2000).

\eject

\section{The $M_\bullet$ --
         $M_{B,{\rm\char'142\char'165\char'154\char'147\char'145}}$ 
         Correlation. II. Bulges Versus Pseudobulges} 

    So far, I have discussed elliptical galaxies and the bulges of disk galaxies
as if they were equivalent.  In terms of BH content, they are indistinguishable:
they are consistent with the same $M_\bullet$ -- $M_{B,\rm bulge}$ and
$M_\bullet$ -- $\sigma_e$ correlations.  But a variety of observational and
theoretical results show that there are two different kinds of high-density
central components in disk galaxies.  Both have steep surface brightness
profiles.  But, while classical bulges in (mostly) early-type galaxies are like
little ellipticals living in the middle of a disk, the ``pseudobulges'' of
(mostly) late-type galaxies are physically unrelated to ellipticals.  

      Pseudobulges are reviewed in Kormendy (1993b) and in Combes (2000).
Observational evidence for disklike dynamics includes (i) velocity dispersions
$\sigma$ that are smaller than those predicted by the Faber-Jackson (1976)
$\sigma$ -- $M_B$ correlation, (ii) rapid rotation $V$ that implies
$V/\sigma$ values well above the ``oblate line'' describing rotationally
flattened, isotropic spheroids in the $V/\sigma$ -- ellipticity diagram, and
(iii) spiral structure that dominates the pseudobulge part of the galaxy.  These
observations and $n$-body simulations imply that high-density central disks can
form out of disk gas that is transported toward the center by bars and oval
distortions.  These heat themselves, for example, by the scattering of
stars off of bars (Pfenniger \& Norman 1990).
Kormendy (1993b) concludes that most early-type galaxies contain traditional
bulges, that later-type galaxies tend to contain pseudobulges, and that only
pseudobulges are seen in Sc -- Sm galaxies.

    Andredakis \& Sanders (1994), Andredakis, Peletier, \& Balcells (1995), and Courteau, de Jong, \& Broeils (1996) show that the ``bulges'' of many late-type
galaxies have nearly exponential surface brightness profiles.  It is likely that
exponential profiles are a signature of pseudobulges, especially since blue
colors imply that they are younger than classical bulges (Balcells \& Peletier
1994).  

    HST observations strengthen the evidence for pseudobulges.  Carollo et al.\
(1997, 1998a, b) find that many bulges have disky properties, including young
stellar populations, spiral structure, central bars, and exponential brightness
profiles.  It seems safe to say that no-one who saw these would suggest that 
they are mini-ellipticals living in the middle of a disk.  Rather, a
morphologist who saw these structures would assign late -- even Im -- Hubble
types.  To be sure, Peletier et al.\ (2000) find that bulges of early-type
galaxies generally have red colors: they are old.  True bulges that are similar
to elliptical galaxies do exist; M{\thinspace}31 and NGC 4594 contain examples.
But the lesson from the Carollo papers is that pseudobulges are more important
than we expected.  Like Kormendy (1993b) and Courteau et al.\ (1996), Carollo
and collaborators argue that these are not real bulges but instead are formed
via gas inflow in disks.

    So there is growing evidence that the ``bulges'' whose luminosities I
used in the $M_\bullet$ -- $M_{B,\rm bulge}$ correlation diagrams are two 
different kinds of objects.  Classical bulges are thought to form like small
ellipticals, via a dissipative collapse, possibly triggered by a merger.  Pseudobulges are thought to form by secular evolution of the disk.  Material
flows inward in both cases.  Potentially, both formation processes may include
BH feeding.  One way to explore this is to ask whether bulges and pseudobulges
have the same BH content. 

\vfill\eject 

\centerline{\null} \vfill
      
\vskip 10pt

\includegraphics{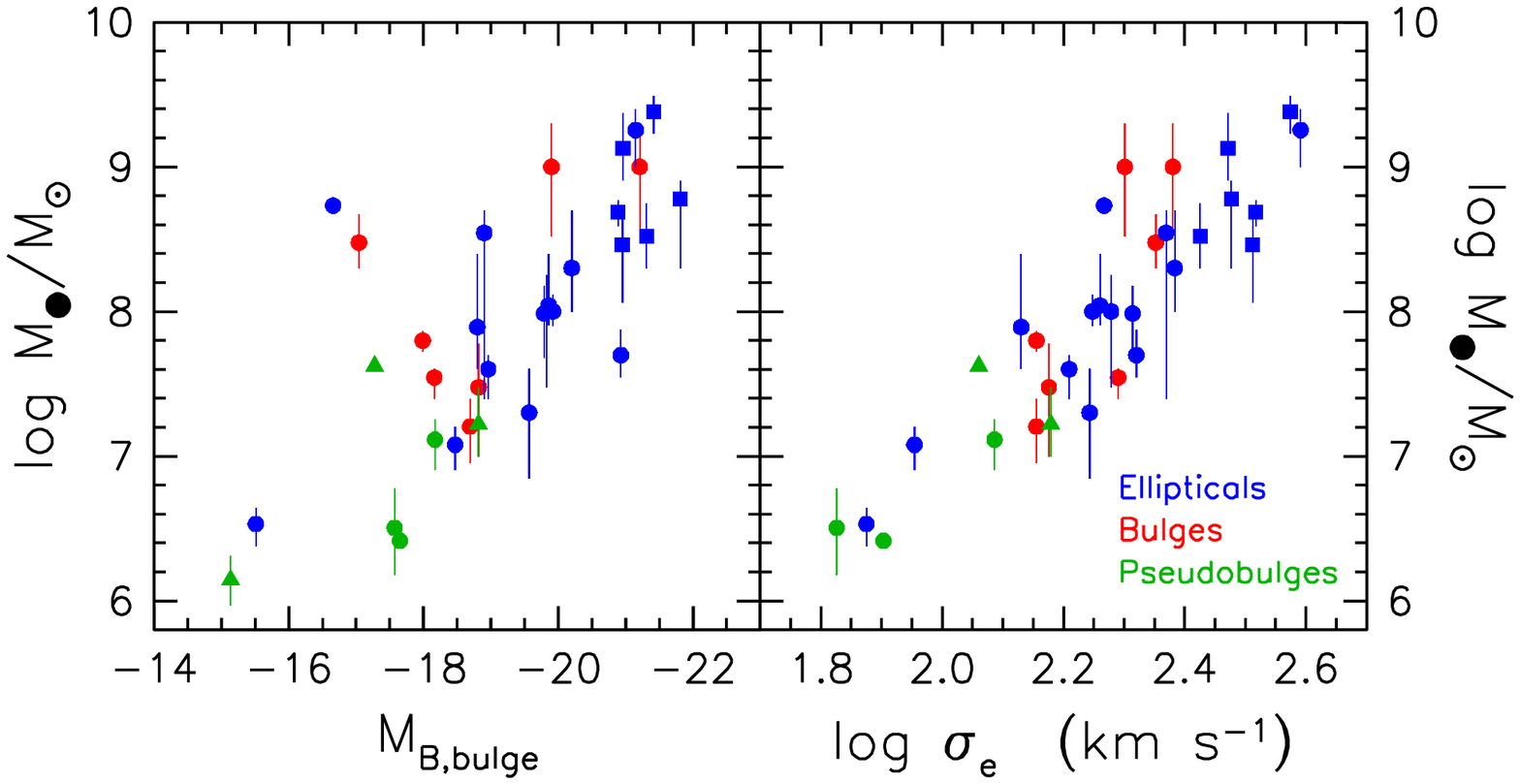}
\noindent Figure 4.{\hskip 1.5em}The $M_\bullet$ -- $M_{B,\rm bulge}$ (left)
and $M_\bullet$ -- $\sigma_e$ (right) correlations for elliptical galaxies
(blue symbols), traditional bulges (red) and pseudobulges (green).   

\vskip 10pt

     This question is addressed in Figure 4.  It is plausible that pseudobulges
have low luminosities, because they are made from disks.  But for their low
luminosities, they have normal BH masses.  The indentification of pseudobulges
is still somewhat uncertain, and there are not many in the sample.  Therefore
the above result needs to be checked.  However, it is consistent with the
hypothesis that (pseudo)bulge formation and BH feeding are closely connected.
Present data do not show any dependence of $M_\bullet$ on the details of whether
BH feeding happens rapidly during a collapse or slowly via secular evolution of
the disk.

      If disks contain only small BHs while the pseudobulges that form from
disks contain standard BHs with 0.2\thinspace\% of the pseudobulge mass, then
I conclude that the BHs must have grown to their present masses during
pseudobulge formation.  This is the second argument that BHs and (pseudo)bulges
grew together.

      The smallest BHs provide an argument that most BH growth did not happen
after bulge formation.  Some pseudobulges are still forming now; there is little
time after bulge formation.  Also, these objects do not contain fuel in the form
of x-ray gas.  And galaxies like M{\thinspace}32 contain little gas of any sort
for late accretion.

\vskip -20pt \null

\section{Conclusion}

Galaxy formation is notoriously complicated; any conclusions that we reach
now are less secure than the observational results discussed in \hbox{\S\S~2 --
4.}  However: {\it Observations increasingly suggest that the major
events that form a bulge and the major growth phases of its BH -- when 
it shone as an AGN -- were the same events.}  The likely formation process is
a series of dissipative mergers that fuel both starbursts and AGN activity
(Sanders et al.\ 1988a, b).

\acknowledgments

      It is a pleasure to thank my Nuker collaborators and especially Karl
Gebhardt and Douglas Richstone for helpful discussions and for permission to 
use our BH detection results before publication.  I am also most grateful to
Gary Bower, Richard Green, Mary Beth Kaiser, and Charles Nelson for 
communicating STIS team BH detections before publication.

\eject

\end{document}